\documentclass[superscriptaddress,pra]{revtex4}

\usepackage[latin1]{inputenc}
\usepackage{graphicx}
\usepackage{amsmath}
\usepackage{amssymb}
\usepackage{dsfont}
\usepackage{graphicx}

\makeatletter
\def\erf{\mathop{\operator@font erf}\nolimits}
\makeatother

\newcommand\be{\begin{equation}}
\newcommand\ee{\end{equation}}

\begin{document}

\title{Non-Markovian interaction of many fields}
\author{R. Mar-Sarao and H. Moya-Cessa}
\affiliation{INAOE, Apdo. Postal 51 y 216, 72000, Puebla, Pue.,
Mexico}

\begin{abstract}
We study the interaction between several fields initially in
coherent states. The solution allows us to explain why coherent
states remain coherent states when subject to non-Markovian
dissipation. We first study the interaction between two fields and
show that this is the building block of the total interaction. We
give a completely algebraic solution of this system.
\end{abstract}
 \maketitle
%\ocis{(270.0270)   Quantum optics, (270.5290)   Photon statistics} % REPLACE WITH CORRECT OCIS CODES FOR YOUR ARTICLE

%%%%%%%%%%%%%%%%%%%%%%% References %%%%%%%%%%%%%%%%%%%%%%%%%

\section{Introduction}
It is well known that a coherent state subject to dissipation
keeps its form during the dynamics. This is, given the master
equation for a field in a lossy cavity at zero temperature
\begin{equation}
\frac{d\rho}{dt}=\gamma(a\rho a^{\dagger}-a^{\dagger}a\rho-\rho
a^{\dagger}a)
\end{equation}
with $a$ ($a^{\dagger}$) the annihilation (creation) operator for
the cavity mode, $\gamma$ the decay constant and $\rho$ the
density matrix, if the initial state of the cavity field is a
coherent state $|\alpha\rangle$, then the dynamics shows that it
will decay in time as $|\alpha e^{-\gamma t}\rangle$ (see for
instance \cite{Moya}. One possible answer about why the coherent
states preserve its form during decay is the fact that coherent
states are eigenstates of the annihilation operator, however this
argument does not hold for a dissipative  two-photon process
\cite{Gilles}
\begin{equation}
\frac{d\rho}{dt}=\gamma(a^2\rho a^{\dagger 2}-a^{\dagger
2}a^2\rho-\rho a^{\dagger 2}a^2)
\end{equation}
even though coherent states are also eigenstates of the
annihilation operator squared (so do even and odd coherent states
\cite{Buzek}).

Both equations above are obtained using Born-Markov approximations
\cite{Gilles,walls}. In the case in which such approximations are
not used, i.e. when the interaction between a harmonic oscillator
and a set of harmonic oscillators (the environment) is considered,
is not clear how a coherent state decays. Here we will try to
answer this question. First we will consider the interaction
between two fields, to later generalize the result to a field
interacting with many.
\section{Two fields interacting}
Consider the Hamiltonian of two interacting fields (we set
$\hbar=1$)
\begin{equation}
H=\omega_a a^{\dagger}a +\omega_b
b^{\dagger}b+\lambda(a^{\dagger}b+b^{\dagger}a), \label{Delta}
\end{equation}%
a system like this may be produced by the interaction of two
quantized fields with a two-level atom \cite{Villas} by
transforming to the interaction picture, i.e. getting rid off the
free Hamiltonians, we obtain
\begin{equation}
H_I=\Delta a^{\dagger}a +\lambda(a^{\dagger}b+b^{\dagger}a),
\label{Delta}
\end{equation}%
with $\Delta=\omega_{a}-\omega_{b}$, the detuning. It is useful to
define normal-mode operators by \cite{Dutra}
\begin{equation}
{A_1}=\delta a + \gamma b, \qquad A_2=\gamma a - \delta b,
\label{transformation}
\end{equation}%
with
\begin{equation}
\delta= \frac{2\lambda}{\sqrt{2\Omega(\Omega-\Delta)}}, \qquad
\gamma=\sqrt{\frac{\Omega-\Delta}{2\Omega}}
\end{equation}%
with $\Omega=\sqrt{\Delta^2+4\lambda^2}$ the Rabi frequency.
${A_1}$ and $A_2$ are annihilation operators just like $a$ and $b$
and obey the commutation relations
\begin{equation}
[{A_1},{A_1}^{\dagger}]=[A_2,A_2^{\dagger}]=1,
\end{equation}%
moreover, the normal-mode operators commute with each other
\begin{equation}
[{A_1},A_2]=[{A_1},A_2^{\dagger}]=0.
\end{equation}
In terms of these operator the Hamiltonian (\ref{Delta}) becomes
\begin{equation}
H_I=\mu_{1} A_1^{\dagger}{A_1} +\mu_{2} {A}_2^{\dagger}{A}_2,
\label{diag}
\end{equation} with $\mu_{1,2}= (\Delta \pm \Omega)/2$.
Up to here, we have translated the problem of solving Hamiltonian
(1) into the problem of obtaining the initial states, for the
"bare" modes  in the initial states for the normal modes. In order
to have a way of transforming states from one basis to the other,
we note that the vacuum states in both systems
$|0\rangle_a|0\rangle_b$ and $|0\rangle_{A_1}|0\rangle_{A_2}$
differ only for a phase \cite{Dutra}. First note that
\begin{equation}
A_1|0\rangle_a|0\rangle_b=0,
\end{equation}
and in a similar way it may be seen the other normal-mode
annihilation operator, $A_2$, has the same effect. Choosing the
phase so that
\begin{equation}
|0\rangle_a|0\rangle_b=|0\rangle_{A_1}|0\rangle_{A_2}.
\end{equation}
If we consider coherent states as initial states for the
interaction, we obtain the evolved wavefunction
\begin{eqnarray} \label{evolv}
\nonumber |\psi(t)\rangle&=&e^{-it(\mu_{1} A_1^{\dagger}{A_1}
+\mu_{2}
{A}_2^{\dagger}{A}_2)}D_a(\alpha)D_b(\beta)|0\rangle_a|0\rangle_b,\\
&=& e^{-it(\mu_{1} A_1^{\dagger}{A_1} +\mu_{2}
{A}_2^{\dagger}{A}_2)}D_a(\alpha)D_b(\beta)|0\rangle_{A_1}|0\rangle_{A_2}
\end{eqnarray}
where the $D_c(\epsilon)=\exp(\epsilon c^{\dagger}-\epsilon^* c)$
is the Glauber displacement operators \cite{Glauber}. From
(\ref{transformation}) we can write the operators $a$ and $b$ in
terms of the operator $A_1$ and $A_2$ (\ref{evolv}) as
\begin{eqnarray}
|\psi(t)\rangle=e^{-it(\mu_{1} A_1^{\dagger}{A_1} +\mu_{2}
{A}_2^{\dagger}{A}_2)}D_{A_1}(\alpha\delta+\beta\gamma)D_{A_2}(\alpha\gamma-\beta\delta)|0\rangle_{A_1}|0\rangle_{A_2}.
\end{eqnarray}
Passing the exponential in the above equation to the right and
applying it to the vacuum states we obtain
\begin{eqnarray} \nonumber
|\psi(t)\rangle&=&D_{A_1}([\alpha\delta+\beta\gamma]e^{-i\mu_1t})D_{A_2}
([\alpha\gamma-\beta\delta]e^{-i\mu_2t})|0\rangle_{A_1}|0\rangle_{A_2}\\
&=&
|[\alpha\delta+\beta\gamma]e^{-i\mu_1t}\rangle_{A_1}|[\alpha\gamma-\beta\delta]e^{-i\mu_2t}\rangle_{A_2}.
\label{14}
\end{eqnarray}
Equation (\ref{14}) shows that in the new basis, coherent states
remain coherent during evolution. By transforming back to the
original basis we obtain
\begin{eqnarray}
|\psi(t)\rangle =
|\delta[\alpha\delta+\beta\gamma]e^{-i\mu_1t}+\gamma[\alpha\gamma-\beta\delta]e^{-i\mu_2t}\rangle_{a}
|\gamma[\alpha\delta+\beta\gamma]e^{-i\mu_1t}-\delta[\alpha\gamma-\beta\delta]e^{-i\mu_2t}\rangle_{b},
\label{15}
\end{eqnarray}
i.e. coherent states remain coherent during evolution. This will
be used next Section as the building block for the interaction of
many modes. In obtaining (\ref{14}) and (\ref{14}), we have used
the following property
\begin{equation}
D_c(\epsilon_1)D_c(\epsilon_2)=D_c(\epsilon_1+\epsilon_1)e^{\frac{1}{2}(\epsilon_1\epsilon_2^*-\epsilon_1^*\epsilon_2)}.
\end{equation}
\section{Generalization to $n$ modes}
Consider the Hamiltonian of the interaction of $k$ fields
\begin{equation}
\hat{H}=\sum_{j}^n\omega _{j}\hat{n}_{j}+
\sum_{j\ne i}^n\lambda _{ij}\left( \hat{a}_{i}^{\dagger }\hat{a}%
_{j}+\hat{a}_{i}\hat{a}_{j}^{\dagger }\right) . \label{1a}
\end{equation}
From the Hamiltonian above, we can produce the following matrix
\begin{equation}
M=\left(
\begin{array}{cccccc}
\omega _{1} & \lambda _{21} & . & . & . & \lambda _{k1} \\
\lambda _{12} & \omega _{2} & . & . & . & \lambda _{k2} \\
\lambda _{13} & \lambda _{23} & . & . & . & \lambda _{k3} \\
. & . & . & . & . & . \\
. & . & . & . & . & . \\
\lambda _{1k} & \lambda _{2k} & . & . & . & \omega _{k}%
\end{array}%
\right) .  \label{11a}
\end{equation}%
 We can
rewrite the Hamiltonian in the form (\ref{diag})
\begin{equation}
\hat{H}=\sum_{m}^n\mu _{m}\hat{A}_{m}^{\dagger }\hat{A}_{m},
\label{2a}
\end{equation}%
such that
\begin{equation}
\left[ \hat{A}_{k},\hat{A}_{m}^{\dagger }\right] =0, \label{3a}
\end{equation}%
where we have defined the normal-mode operators $\hat{A}_{k}$ as
\begin{equation}
\hat{A}_{k}=\sum_{i=1}^nr_{ki}\hat{a}_{i}, \label{4a}
\end{equation}%
with $r_{ki}$ a real number.

Equation (\ref{3a}) implies that%
\begin{equation}
\left[ \hat{A}_{k},\hat{A}_{m}^{\dagger }\right] =\sum_{i,j=0}^nr_{ki}r_{mj}\left[ \hat{a}_{i},\hat{a}_{j}^{\dagger }%
\right] =\sum\limits_{i}^nr_{ki}r_{mi}=0.  \label{5a}
\end{equation}%
By defining the vector
\begin{equation}
\vec{r}_{k}=\left( r_{k1},r_{k2},...,r_{kn}\right) , \label{6a}
\end{equation}%
equation (\ref{5a}) takes the form $\vec{r}_n\cdot\vec{r}_m=0$,
i.e. they are orthogonal, we will consider them also normalized,
$\vec{r}_k\cdot\vec{r}_k=1$. With these vectors we can form the
matrix
\begin{equation}
{R}=\left(
\begin{array}{cccccc}
r_{11} & r_{21} & . & . & . & r_{n1} \\
r_{12} & r_{22} & . &  & . & r_{n2} \\
. & . & . & . & . & . \\
. & . & . & . & . & . \\
. & . & . & . & . & . \\
r_{1n} & r_{2n} & . & . & . & r_{nn}%
\end{array}%
\right) ,  \label{7a}
\end{equation}%
If we combine equations (\ref{1a}), (\ref{2a}) and (\ref{4a}) we
obtain the system of equations
\begin{equation}
\sum_{m}\mu _{m}r_{mi}^{2}=\omega _{i},  \label{9a}
\end{equation}%
\begin{equation}
\sum_{m}\mu _{m}r_{mi}r_{mj}=\lambda _{ij},. \label{10a}
\end{equation}%
that  may be re-expressed in the compact form
\begin{equation}
{R}{D}{R}^{\dagger}={{M}}=\left(
\begin{array}{cccccc}
\omega _{1} & \lambda _{21} & . & . & . & \lambda _{n1} \\
\lambda _{12} & \omega _{2} & . & . & . & \lambda _{n2} \\
\lambda _{13} & \lambda _{23} & . & . & . & \lambda _{n3} \\
. & . & . & . & . & . \\
. & . & . & . & . & . \\
\lambda _{1n} & \lambda _{2n} & . & . & . & \omega _{n}%
\end{array}%
\right) ,  \label{11a}
\end{equation}%
with
\begin{equation}
{{D}}=\left(
\begin{array}{cccccc}
\mu _{1} & 0 & . & . & . & 0 \\
0 & \mu _{2} & . & . & . & 0 \\
0 & 0 & . & . & . & . \\
. & . & . & . & . & . \\
. & . & . & . & . & . \\
0 & 0 & . & . & . & \mu _{n}%
\end{array}%
\right) ,  \label{8a}
\end{equation}%
i.e. ${D}$ is a diagonal matrix whose elements are the eigenvalues
of the matrix ${M}$, defined from the Hamiltonian. The matrix
${R}$ is therefore ${M}$'s eigenvectors matrix. The solution to
the Schr\"odinger equation subject to the Hamiltonian (\ref{1a})
with all the modes initially in coherent states,
$|\psi(0)\rangle=|\alpha_1\rangle_1 |\alpha_2\rangle_2 ...
|\alpha_n\rangle_n $, is simply the direct product of coherent
states
\begin{equation}
|\psi(t)\rangle=|\vec{r}_1 \cdot\vec{\beta}(t)\rangle_1
|\vec{r}_2\cdot\vec{\beta}(t)\rangle_2 ...
|\vec{r}_n\cdot\vec{\beta}(t)\rangle_n
\end{equation}%
with $\vec{\beta}(t)=(\vec{r}_1\cdot\vec{\alpha}e^{-i\mu_1t},
\vec{r}_2\cdot\vec{\alpha}e^{-i\mu_2t}, ...,
\vec{r}_n\cdot\vec{\alpha}e^{-i\mu_nt})$ and the vector
$\vec{\alpha}=({\alpha}_1, {\alpha}_2, ... ,{\alpha_n})$ is
composed by the coherent amplitudes of the initial wave function.
Up to here we have shown that the interaction of several modes
initially in coherent states does not change the form of those
states (remain coherent), but modifies their amplitude. If we
choose the interaction constants to be $\lambda_{1j}\ne 0$ for
$1\ne j$ and the rest as zero, we are dealing with the interaction
between one field and $n-1$ fields. If $n\rightarrow\infty$ and
the amplitudes $\alpha_j$ are zero for $j>1$, we deal with the
interaction of one field with $n-1$ one of them in a coherent
state with amplitude $\alpha_1$ and the rest in the vacuum.
Therefore, the most likely situation we have is the coherent state
decaying towards the vacuum while keeping its coherent form.

\section{Conclusions} We have shown that a system of $n$
interacting harmonic oscillators initially in coherent states,
remain coherent during the interaction. In particular, if one
considers one field (harmonic oscillator) interacting with many
fields (harmonic oscillators), i.e. consider only $\lambda_{1j}\ne
0$ and  $\lambda_{j1}\ne 0$ for $j>2$, and all the others to be
zero, we can model non-Markovian system-reservoir interaction. If
we consider the system to be in a coherent state and all the
others fields that form the environment in a vacuum state (this is
also in coherent states with zero amplitude), after evolution, the
amplitude of the coherent state will diminish, as one photon will
go to another mode, keeping its coherent nature. If the number of
modes that form the environment is very large, an event of the
photon going back to the system is quite unlikely. Therefore the
next probable event is precisely the loss of another photon by the
system, etc. until it arrives to a state close to the vacuum. In
case the number of modes interacting with the system is infinite,
then the vacuum would be the final state of the system. In other
words, the total system perform the following transition
\begin{equation}
|\alpha\rangle_1|0\rangle_2 \dots |0\rangle_n \rightarrow
|\delta_1\rangle_1|\delta_2\rangle_2 \dots |\delta_2\rangle_n,
\end{equation}
where the coherent amplitudes,$ \delta_k\rightarrow 0$, as
$n\rightarrow \infty$.

In conclusion we have given a complete algebraic solution to the
problem of $n$ interacting harmonic oscillators, without
Born-Markov approximations.

\end{document}